\documentstyle[graphicx,epsf]{mn}

\newcommand{\ls}
 {\mathrel{\hbox{\rlap{\hbox{\lower4pt\hbox{$\sim$}}}\hbox{$<$}}}}
\newcommand{\gs}       
 {\mathrel{\hbox{\rlap{\hbox{\lower4pt\hbox{$\sim$}}}\hbox{$>$}}}}
\newcommand{\degrees}{\hbox{$^\circ$}}

\newcommand{\et}{et al.}
\newcommand{\etal}{et al. }
\newcommand{\exosat}{{\it EXOSAT }}
\newcommand{\rosat}{{\it ROSAT }}

\newcommand{\asca}{{\it ASCA }}
\newcommand{\ginga}{{\it GINGA }}

\title[ASCA X-ray observations of Z~Cam]
	{ASCA X-ray observations of the disk wind in the dwarf nova Z~Camelopardalis }
\author[Darren S.\ Baskill, Peter J.\ Wheatley and Julian P.\ Osborne]
	{D.S.\ Baskill, P.J.\ Wheatley and J.P.\ Osborne\\  
X-Ray Astronomy Group, Department of Physics and Astronomy, Leicester 
	University, Leicester LE1 7RH, U.K.\\
}

\date{Submitted }

\pagerange{\pageref{firstpage}--\pageref{lastpage}}
\pubyear{2001}

\begin{document}

\maketitle

\label{firstpage}

\begin{abstract}

We present \asca observations of the dwarf nova Z~Camelopardalis during outburst and during a transition from quiescence to another outburst.

At the beginning of the transition the X-ray count rate was an order of magnitude higher and the spectrum much harder than during the outburst.  As the transition progressed, the spectrum remained hard as the X-ray flux decreased by a factor of 3, with no spectral softening.

Spectral modelling reveals an optically-thin, high-temperature component (kT$\approx$10\,keV) which dominates the transition observation and is also observed during outburst.  This is expected from material accreting onto the white dwarf surface.  
The outburst spectra require additional emission at lower temperatures, either through an additional discrete temperature component, or a combination of a cooling flow model and an ionised absorber.

Fits to both observations show large amounts of absorption ($N_H=8-9\times10^{21}$cm$^{-2}$), two orders of magnitude greater than the measured interstellar value, and consistent with UV measurements of the outburst.  This suggests that a disk wind is present even in the earliest stages of outburst, possibly before the outburst heating wave has reached the boundary layer.

\end{abstract}


\section{Introduction}

\begin{figure*}
\includegraphics{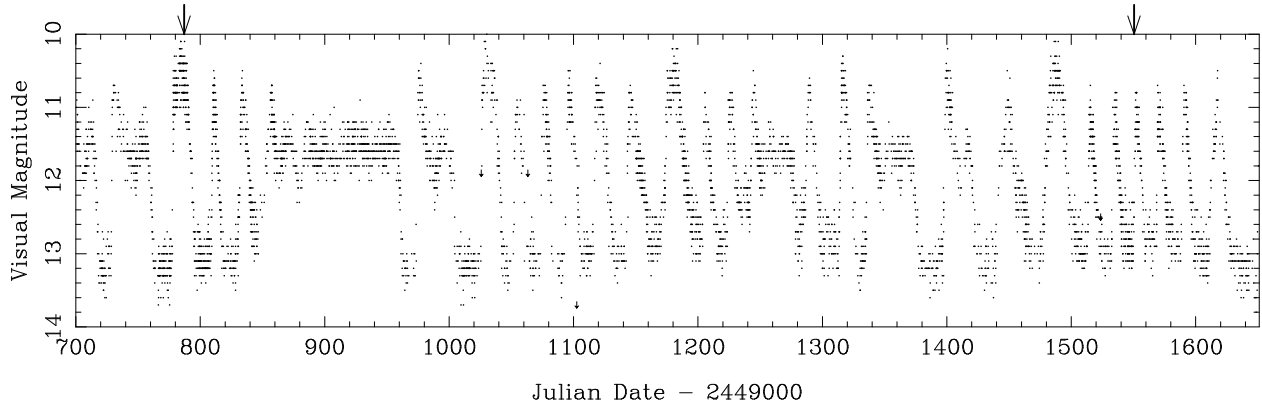}
\caption{Optical light-curve of Z Cam from AAVSO observations from January 1995 until July 
1997.  The times of the \asca observations are shown at JD 2449783/4 and JD 2450551}
\label{fig:longoptical}
\end{figure*}

Eclipse observations of dwarf novae during quiescence show that the
X-ray emitting region is centred on the white dwarf, within a size that
is comparable to that of the white dwarf (e.g. HT~Cas, Mukai \et, 1997).  
The emission is believed to originate from material  in a boundary layer slowing from a Keplerian velocity in the disk, to settle onto the white dwarf surface.  

  Observations of some systems have shown hard X-ray emission dominating during quiescence and an extreme-ultraviolet component dominating during outburst, when the hard X-ray component is suppressed (Patterson \& Raymond, 1985a~\&~1985b, SS~Cyg by Cordova \etal 1984, U~Gem by Cordova \etal 1980 and Mason \etal 1988, VW~Hyi by Pringle \etal 1987).
The X-ray flux recovers only at the very end of the optical outburst (Wheatley \et, 1996a).  

During quiescence both the accretion rate and the boundary layer density are low, cooling occurs inefficiently, and so the temperature remains high (Pringle \& Savonije, 1979).
  During outburst the mass 
accretion rate increases, the boundary layer becomes optically thick and efficient at cooling, thus the emission of hard X-rays is reduced (Pringle 1977).  Such an optically thick boundary layer cools to an effective blackbody temperature of 10$^5$K (10\,eV) in the extreme ultra-violet (Patterson \& Raymond, 1985b).

  X-ray observations with the \rosat satellite confirm that this picture also applies to Z~Cam during outburst (Wheatley \et, 1996b).  
However, these \rosat data suffer from the low exposure of the 
all-sky survey, as well as the intrinsic limitations of a soft X-ray
proportional counter in the study of hard spectra.

Done \& Osborne (1997) have made a detailed analysis of the X-ray spectrum
of the dwarf nova SS Cyg using the \ginga and \asca satellites.  They find that physically plausible spectral fits require models for gas cooling onto the white dwarf surface, with absorption by a photo-ionised medium.  They find that a greater proportion of cool gas and an increase in X-ray reflection is required in outburst than in quiescence.

In this paper we present \asca X-ray spectroscopic observations 
which caught the dwarf nova Z~Cam in outburst in 1995 and during a transition from quiescence to outburst in 1997 (see Fig.~\ref{fig:longoptical}).

Knigge \etal (1997) observed the same outburst with the Hopkins Ultraviolet Telescope on the Astro-2 mission.  
They apply an accretion disk wind model to fit absorption features in the UV spectrum of Z~Cam, and conclude that a dense, slow-moving disk wind transition region acts as an absorbing medium with a temperature of few$\times10^{4}$.

  Modelling of disk winds has been carried out by Proga \etal (1998) who find that radiation driven
winds are intrinsically unsteady, with large density and velocity variations.  The dense low-velocity flow component of the wind is confined to angles below 45\degrees of the equatorial plane.  Z~Cam has an inclination of 57$\pm$11\degrees (Shafter 1983) and so the line of sight to Z~Cam should pass through this zone.

\section{Observations}
Z~Cam has been observed in both outburst and during the optical rise to another outburst using the Japanese \asca satellite.
\asca has four X-ray telescopes and detectors: two X-ray CCD cameras (Solid-state
Imaging Spectrometers, or SIS) and two gas scintillation imaging
proportional counters (Gas Imaging Spectrometers, or GIS) (Tanaka,
Inoue \& Holt 1994).  The SIS detectors have an energy range of 0.4-12\,keV whilst the GIS have an energy range of 0.7-15\,keV.  

\begin{figure*}
\includegraphics{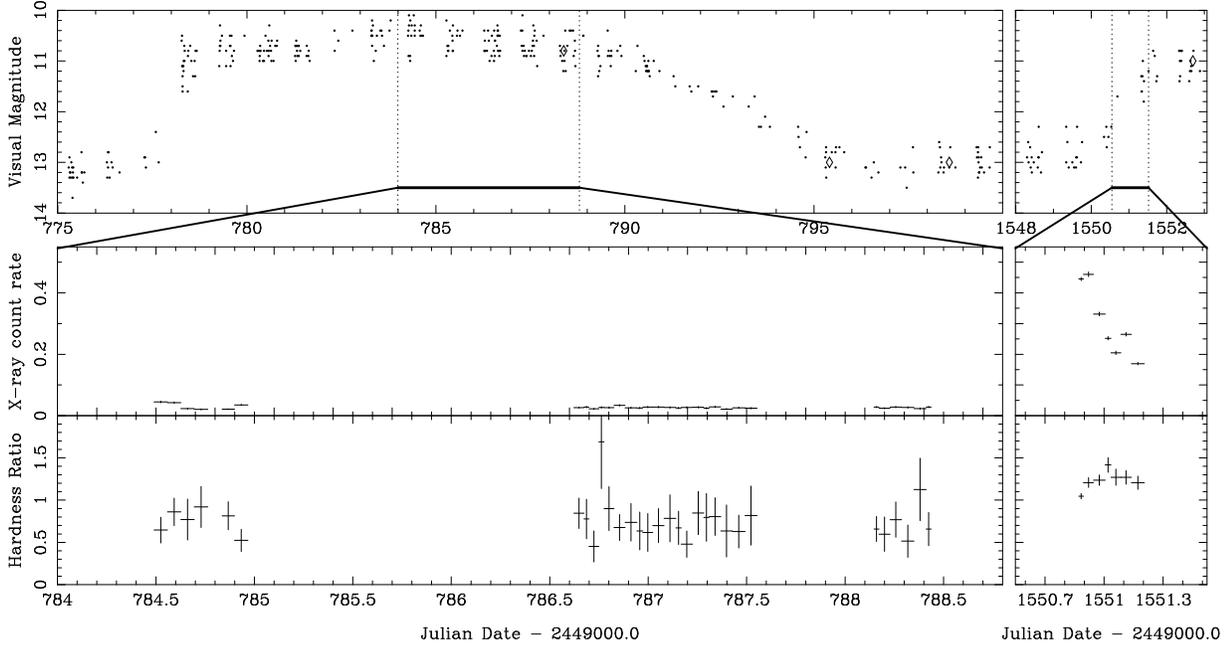}
\caption{Z Cam in outburst (left) and transition
(right) showing the AAVSO visual light-curve (top), the ASCA X-ray
light-curve from all four instruments (middle) and the X-ray hardness ratio ($>$2.4\,keV light-curve divided by
the $<$2.4\,keV light-curve) from both GIS instruments.}
\label{fig:opticalxray}
\end{figure*}

  As shown in Figs~\ref{fig:longoptical}~\&~\ref{fig:opticalxray}, Z~Cam was observed 
on two occasions:  once during outburst between 8 March to 12 March, 1995 
(JD 2449783-2449788), and once during a transition as Z~Cam optically brightened to another outburst state on 12 April, 1997 (JD 2450550).  
  To maximise the amount of data, and to avoid unnecessary gaps in the light curve, loose screening criteria have been applied compared to the standard REV2 screening.  We reject data from all instruments in both observations when the telescope has deviated 
more than 0.01\degrees away from the source.
  The elevation angle between the Earth's limb and satellite pointing
direction has been limited to greater than 10\degrees with the SIS and
5\degrees with the GIS, the bright Earth elevation angle is limited to greater
than 15\degrees and the radiation background monitor is restricted to
less than 250 counts per second.  The minimum value of the cut-off
rigidity of the geomagnetic field throughout both observations is approximately 4 GeV/c, which is high enough not to require an additional screening criterion.  Time filters have been applied to cut out the few remaining periods of very high background during the transitional observation, although this amounts to less than 300 seconds in all instruments.  
  Circular source extraction regions have been used of 6 arcmin diameter in the GIS and 4 arcmin in the SIS.  The background has been taken from the remaining area of the same chip in the SIS, and an annulus of outer radius 18 arcmin around the source for the GIS. 

  The outburst observation exposure times after screening are 80\,ks \& 76\,ks for SIS0 \& SIS1
and 94\,ks for each GIS.  For the transition observation, the exposure times are 18~ks \& 22~ks for the 
two SIS, and 22~ks \& 23~ks for the two GIS.
The response matrices for each SIS have been generated using the FTOOL SISRMG, and for the 
GIS the standard response matrices (version 0.8, November 1994) have been used. 
After this loose screening, the two SIS instruments and two GIS instruments were 
separately combined in both observations. 
The outburst spectra have been grouped to a minimum of 40 counts per bin and the transition spectra have been grouped to at least 20 counts per bin, allowing $\chi^{2}$ fitting. 

\begin{figure*}
\includegraphics{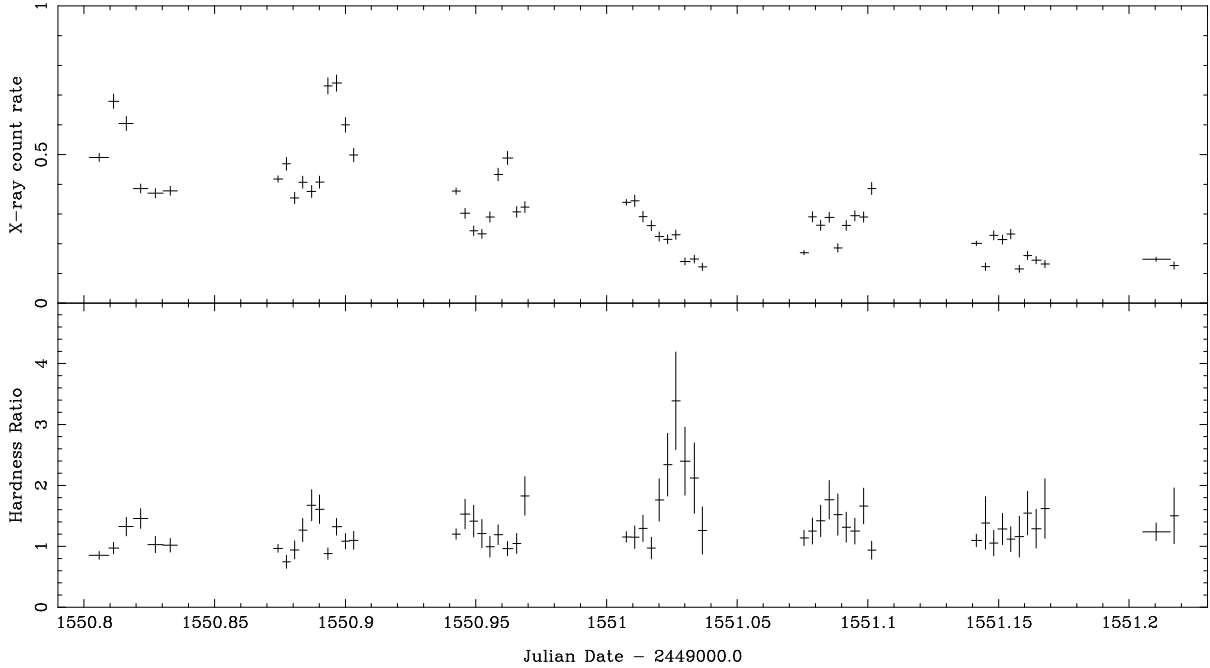}
\caption{The \asca X-ray light-curve from all four instruments (top) and the X-ray hardness ratio ($>$2.4\,keV light-curve divided by
the $<$2.4\,keV light-curve) from both GIS instruments (bottom) during the transition observation.}
\label{fig:transcurve}
\end{figure*}

\section{Light curves}

Figure~\ref{fig:opticalxray} shows the optical light curve, the X-ray light curve and the X-ray hardness ratio during the two \asca observations.  The top panel shows the AAVSO optical observations, and the bottom two panels show the \asca data over a one sixth smaller time range.
  The hardness ratio shows the plot of the hard light curve ($>$2.4\,keV) over the soft light curve ($<$2.4\,keV).  
Each bin in Fig.~\ref{fig:opticalxray} represents one \asca orbit, with each bin defined by gaps in the data due to Earth occultations and regions of high background.

  The optical light curves show that the \asca observations were taken when Z~Cam was in an outburst state, and during an optical transition to outburst.  
  Throughout the outburst observation the X-ray count rate was constant, after small initial variations.  At the onset of the transition, the X-ray count rate was twenty times the outburst observation count rate.  During the transition the X-ray count rate fell dramatically by a factor of 3.

The hardness ratio remained constant throughout the outburst, and was softer than during the optical transition.  As the X-ray count rate fell during the
transition to outburst, the hardness ratio might have been expected to soften to the level of the earlier outburst observation.  However, this is not the case.  The hardness ratio appears to increase until midway through the transition, remaining harder than the outburst observation throughout. 

  Fig.~\ref{fig:transcurve} shows the X-ray light curve and hardness ratio of the transitional observation in more detail.  The light curve shows a decreasing count rate with irregular peaks.  Apart from the large peak in hardness ratio, there appears to be little correlation between the count rate and hardness ratio, with the hardness ratio remaining approximately constant throughout. 

\section{Spectral Analysis}

  We selected two pairs of spectra for detailed analysis.  These are the SIS \& GIS spectra from throughout the outburst and from the entire transition observation.

  All the  spectra have been fitted with several standard models as used in the XSPEC spectral fitting package, version 11.0 (Arnaud 1996).
  Throughout the modelling, both SIS and GIS have been fitted simultaneously, with a energy-independent multiplicative factor applied between the SIS and GIS instruments as a free parameter.

\begin{figure*}
\includegraphics{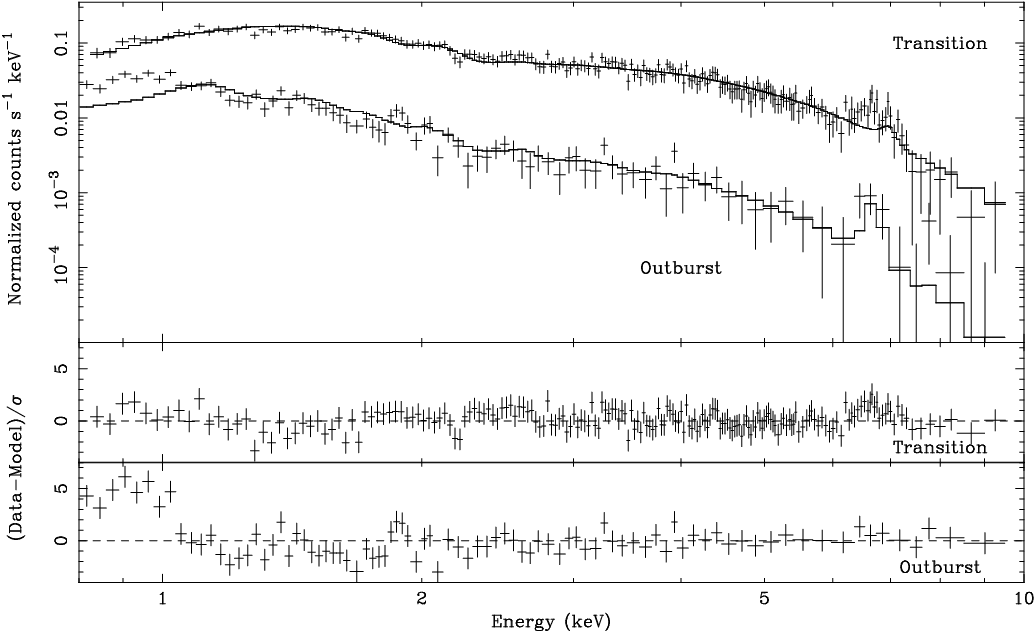}
\caption{The \asca spectra of the outburst and transition observations, showing 
the single temperature model fits.  Only the SIS spectra are plotted for clarity.  
The lower panels show the deviation of the data from the model in units of statistical error, $\sigma$.}
\label{fig:1tspectra}
\end{figure*}

\begin{figure*}
\includegraphics{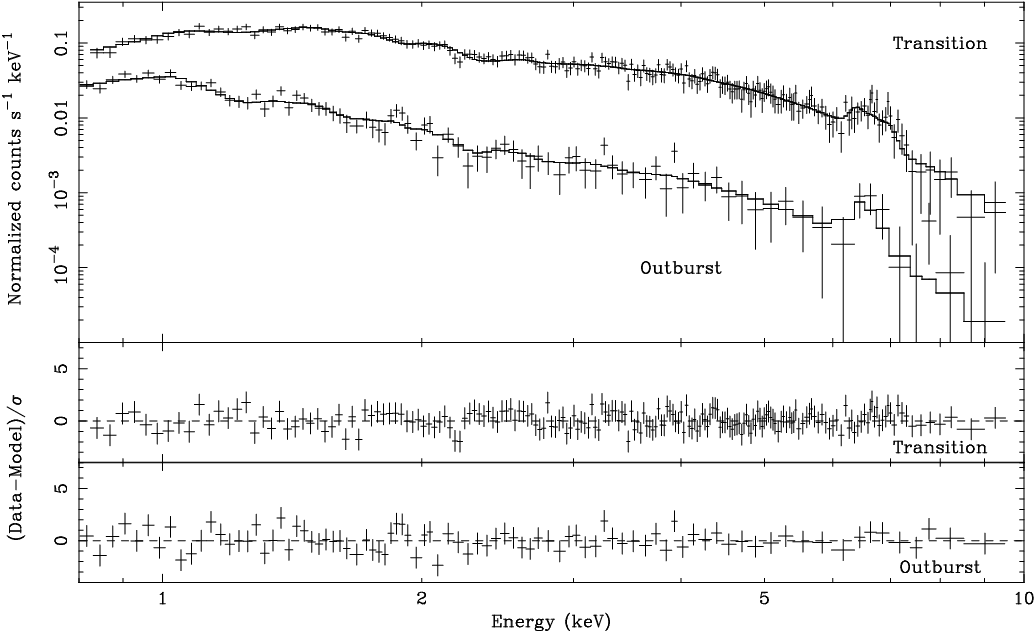}
\caption{The \asca SIS spectra from the outburst and transition 
observations, showing our best-fitting DO97 model (see section~4.2 for details).  The lower panels show the deviation of the data from the model in units of statistical error, $\sigma$.}
\label{fig:spectra}
\end{figure*}

\subsection{Discrete Temperature models}

We started the spectral fitting of Z~Cam using a single temperature {\it mekal} model for emission from a hot diffuse gas (Mewe \etal 1985, Mewe \etal 1986, Liedahl \etal 1995) and 
{\it wabs} to model photo-electric absorption (Morrison \& McCammon, 1983). 
This leads to acceptable fits for the transition spectrum ($\chi^2_\nu=0.77$, 603 degrees of freedom) but poor fits to the outburst spectrum ($\chi^2_\nu=1.72$, 225 d.o.f.).
 The single temperature fits can be seen in Fig.~\ref{fig:1tspectra}, showing excess emission below 1\,keV, especially in the outburst observation.  

In an attempt to model the excess emission below 1\,keV a second {\it mekal} component ($kT\approx0.6$\,keV) was added to the model.  This component only contributes line emission around 1\,keV, predominantly iron L-shell emission, thus resulting in an improved fit. Thus, acceptable fits where obtained to both the outburst and transition spectra (see table~\ref{table:2ts}).

Not only does this two-temperature model fit both spectra, but it is only necessary to allow one parameter to vary between the two states, namely the normalisation of the higher temperature component ($\chi^2_\nu=0.71$, 829 d.o.f., fitting both pairs of spectra simultaneously).  Thus the spectra of Z~Cam can be interpreted as originating from two, single-temperature plasmas, with only the amount of hot gas varying between the two observations.

\begin{table*}
 \centering
  \begin{tabular}{||l|l|c|c||}        
Component & & Transition & Outburst\\ \hline
\multicolumn{2}{l}{Equivalent hydrogen column Absorption (atoms/cm$^{-2}$)} & $11.2^{+1.7}_{-2.4}\times 10^{21}$ & $3.1^{+0.4}_{-0.3}\times 10^{21}$\\ 
Lower temperature: & Temperature (keV) & $0.25^{+0.07}_{-0.05}$ & $0.69^{+0.08}_{-0.08}$\\ 
& Normalisation
& $40.6^{+63.2}_{-29.0}\times 10^{-3}$ 
& $0.32^{+0.61}_{-0.16}\times 10^{-3}$\\ 
& Flux (0.8-10.0 keV, ergs cm$^{-2}$ s$^{-1}$) & $1.3\times10^{-12}$ & $0.3\times10^{-12}$ \\
& Bolometric Luminosity (0.1-100.0 keV, ergs s$^{-1}$) & $0.5\times10^{31}$ & $0.1\times10^{31}$ \\
Higher temperature: & Temperature (keV) & $7.7^{+1.4}_{-1.0}$ & $7.5^{+4.0}_{-1.6}$ \\
& Normalisation & 
$14.3^{+0.8}_{-0.8}\times 10^{-3} $ & 
$0.59^{+0.04}_{-0.05}\times 10^{-3}$ \\ 
& Flux (0.8-10.0 keV, ergs cm$^{-2}$ s$^{-1}$) & $2.0\times10^{-11}$ & $0.1\times10^{-11}$ \\
& Bolometric Luminosity (0.1-100.0 keV, ergs s$^{-1}$) & $9.4\times10^{31}$ & $0.4\times10^{31}$ \\
\multicolumn{2}{l}{$\chi^2$} & 398 & 150 \\ 
\multicolumn{2}{l}{Degrees of freedom} & 601 & 223 \\ 
\multicolumn{2}{l}{$\chi^2_\nu$} & 0.66 & 0.67\\ 
\multicolumn{2}{l}{Total Flux (0.8-10.0 keV, ergs cm$^{-2}$ s$^{-1}$)} & $2.2\times10^{-11}$ & $1.3\times10^{-12}$ \\
\multicolumn{2}{l}{Bolometric Luminosity (0.1-100.0 keV, ergs s$^{-1}$)} & $9.9\times10^{31}$ & $0.6\times10^{31}$\\
\hline
Component & & Transition & Outburst\\ \hline
\multicolumn{2}{l}{Equivalent hydrogen column Absorption (atoms/cm$^{2}$)} & \multicolumn{2}{c}{$4.8^{+0.4}_{-0.4}\times 10^{21}$} \\ 
Lower temperature: & Temperature (keV) & \multicolumn{2}{c}{$0.65^{+0.03}_{-0.05}$} \\ 
& Normalisation & \multicolumn{2}{c}{ $0.59^{+0.09}_{-0.08}\times 10^{-3}$} \\ 
& Flux (0.8-10.0 keV, ergs cm$^{-2}$ s$^{-1}$) & \multicolumn{2}{c}{$3.9\times10^{-13}$} \\ 
& Bolometric Luminosity (0.1-100.0 keV, ergs s$^{-1}$) & \multicolumn{2}{c}{$1.5\times10^{30}$} \\
Higher temperature: & Temperature (keV) & \multicolumn{2}{c}{$13.9^{+4.0}_{-2.1}$} \\
& Normalisation & 
$12.2^{+0.3}_{-0.3}\times 10^{-3}$
& $0.56^{+0.04}_{-0.04}\times 10^{-3}$\\
& Flux (0.8-10.0 keV, ergs cm$^{-2}$ s$^{-1}$) & $21.1\times10^{-12}$ & $0.98\times10^{-12}$\\
& Bolometric Luminosity (0.1-100.0 keV, ergs s$^{-1}$) & $1.2\times10^{32}$ & $5.7\times10^{30}$ \\
\multicolumn{2}{l}{$\chi^2$} & \multicolumn{2}{c}{586} \\ 
\multicolumn{2}{l}{Degrees of freedom}  & \multicolumn{2}{c}{829} \\ 
\multicolumn{2}{l}{$\chi^2_\nu$} & \multicolumn{2}{c}{0.71} \\ 
\multicolumn{2}{l}{Total Flux (0.8-10.0 keV, ergs cm$^{-2}$ s$^{-1}$)} & $21.5\times10^{-12}$ & $1.4\times10^{-12}$ \\
\multicolumn{2}{l}{Bolometric Luminosity (0.1-100.0 keV, ergs s$^{-1}$)} & $1.2\times10^{32}$ & $7.2\times10^{30}$\\
\hline
\end{tabular}
\caption{Results from the two-temperature {\it mekal} models.  The top table shows our best fit to the outburst and transition spectra independently.  The lower table shows our best fit to the outburst and transition spectra simultaneously, with just a single parameter free to vary between the two states.  All errors are to the 90\% confidence level for the single interesting parameter, equivalent to a $\Delta\chi^2$ statistic of 2.706.
Normalisations are as implemented in the XSPEC software, in units of 
$\frac{10^{-14}}{4\pi d^{2}EM}$ where $EM$ is the volume emission measure and $d$ is the distance to the source.   The bolometric luminosity assumes a distance of 170pc (Warner, 1987).}
\label{table:2ts}
\end{table*}

  The two-temperature model fits yield large values for the absorption during both the transition ($N_H=11.2^{+1.7}_{-2.4}\times 10^{21}$ cm$^{-2}$) and the outburst ($N_H=3.1^{+0.4}_{-0.3}\times 10^{21}$cm$^{-2}$). 
 The absorption throughout both the outburst and transition is two orders of magnitudes greater than the value of $N_H=4\times 10^{19}$cm$^{-2}$ determined from an {\it IUE} curve-of-growth study of interstellar absorption lines (C. Mauche, private communication, for method see Mauche, Raymond \& Cordova 1988).  This excess over the interstellar value suggests that a large amount of absorption is occurring close to the X-ray source, even in the earliest stages of outburst.

  Since the excess absorption is local to the system, it is unrealistic to expect the absorbing medium to be neutral.  Therefore, the neutral absorber was replaced with an ionised absorber in the single temperature {\it mekal} model.  Although this again produces an acceptable fit to the transition spectra ($\chi^2_\nu=0.67$, 601 d.o.f.) it cannot reproduce the outburst spectra ($\chi^2_\nu=1.65$, 223 d.o.f.). 

\begin{figure}
\includegraphics{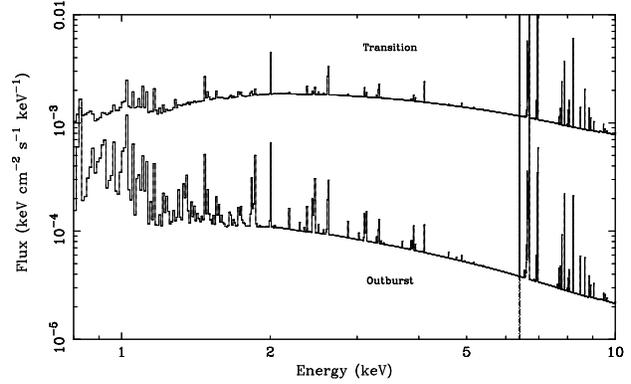}
\caption{A plot of the incident spectrum for the best-fitting DO97 model, during transition (top) and outburst (bottom).}
\label{fig:donemodels}
\end{figure}

\begin{figure}
\includegraphics{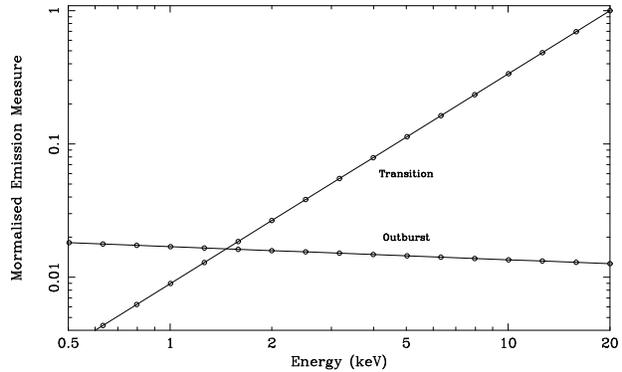}
\caption{A plot of the differential emission measure for the best-fitting DO97 model during transition and outburst, with a maximum temperature of 20\,keV.}
\label{fig:donedem}
\end{figure}

\subsection{Continuous temperature distribution models}

 In the previous section we have shown that the data can be modelled with a optically-thin plasma at two distinct temperatures.  However, the X-rays are expected to originate from only one site, with a continuous range of temperatures representing gas cooling onto the white dwarf.

  In an attempt to fit our spectra with a continuous temperature distribution, we adopted the model of Done \& Osborne (1997), as applied to the \asca quiescent spectrum of SS~Cyg (their {\it pia(plT+g)} model).  The model consists of neutral photo-electric absorption to model interstellar absorption (which we freeze at $N_H=4\times 10^{19}$cm$^{-2}$), and an ionised absorption component is required to model absorption local to the system.
The emission  measure of the X-ray emitting plasma is modelled as a power-law in temperature, $(T/T_{Max})^\alpha$ using the {\it cevmkl} model in XSPEC.  
The \asca observations of Z~Cam cannot constrain the reflection continuum component due to the low number of counts at high energies, and so this has been omitted in our version of the model.  However, a gaussian is included to model the 6.4\,keV fluorescent line.
We shall refer to this model as the DO97 model hereafter.

\begin{table*}
 \centering
  \begin{tabular}{||l|l|c|c||}        
Component & & Transition & Outburst\\ \hline
Neutral Absorber & Absorption n$_H$ (atoms/cm$^2$)& $4\times10^{19}*$ & $4\times10^{19}*$ \\
Partially ionised Absorber & Absorption n$_H$ (atoms/cm$^2$)& $9.0^{+1.6}_{-1.5}\times10^{21}$ & $7.8^{+0.4}_{-0.3}\times10^{21}$\\
& Temperature (K) & $3.6^{+3.6}_{-3.6}\times10^{4}$ & $4.1^{+6.4}_{-4.1}\times10^{4}$ \\
& Ionisation Parameter $\xi$ (L/nR$^2$, see Done \etal 1992) & $17.8^{+92.1}_{-8.2}$ & $8.0^{+67}_{-4.9}$ \\
Continuous temperature emission & power-law index & $1.6^{+0.5}_{-0.3}$ & $-0.1^{+0.1}_{-0.2}$ \\
& T$_{max}$ (keV) & $20*$ & $20*$ \\
& Metal abundance & $1.0*$ & $1.0*$ \\
& Normalisation & $4.0^{+0.9}_{-0.4}\times10^{-2}$ & $5.1^{+1.6}_{-1.4}\times10^{-4}$\\
Gaussian & Energy (keV) & $6.4*$ & $6.4*$ \\
& Equivalent width (eV) & $160^{+50}_{-60}$ & $400^{+120}_{-90}$  \\
\multicolumn{2}{l}{$\chi^2$} & 386 & 150\\ 
\multicolumn{2}{l}{Degrees of freedom} & 600 & 222\\ 
\multicolumn{2}{l}{$\chi^2_\nu$} & 0.64 & 0.68\\ 
\multicolumn{2}{l}{Flux (0.8-10.0 keV, ergs cm$^{-2}$ s$^{-1}$)} & $2.2\times10^{-11}$ & $0.13\times10^{-11}$\\
\multicolumn{2}{l}{Bolometric Luminosity (0.1-100.0 keV, ergs s$^{-1}$)} & $1.3\times10^{32}$ & $9.3\times10^{30}$\\
\hline

 \end{tabular} 
\caption{Our best-fitting DO97 model (see text for details).  All errors are to the 90\% confidence level ($\Delta\chi^2=2.706$).  The bolometric luminosity assumes a distance of 170pc. (*Parameter frozen at this value)}
\label{table:done}
\end{table*}

The results of the modelling can be seen in Table~\ref{table:done} and Figs.~\ref{fig:spectra},~\ref{fig:donemodels},~\ref{fig:donedem}~\&~\ref{fig:nhvsxi}.
  The metal abundances used throughout are those of Anders \& Grevesse (1989), with the abundances frozen at solar values.  Thawing the abundances leads to no improvement for the quiescent observation, and marginal improvement to the outburst observation ($\Delta\chi^2=3.0$). 
  The maximum temperature of the power-law temperature emissivity component is poorly constrained and so is fixed at 20\,keV.

  The DO97 model fits as well as the two-temperature model, with $\chi^2_\nu=0.64$ (600 d.o.f.) during the transition and $\chi^2_\nu=0.68$ (222 d.o.f.) during outburst.  
The residuals in the single temperature fits below 1\,keV (Fig.~\ref{fig:1tspectra}) are modelled in the DO97 model by a combination of increased line emission from cool gas and increased absorption.
  We find that both the ionisation state and absorption column of the absorber remains approximately constant in both the outburst and the transition (see Fig.~\ref{fig:nhvsxi}).
The power-law emissivity-temperature distribution is weighted towards higher temperature components in the transition observation ($\alpha=1.6^{+0.5}_{-0.3}$) as would be expected of a cooling flow, and lower temperature components during the outburst ($\alpha=-0.1^{+0.1}_{-0.2}$) which shows an excess of cool gas (see Figs.~\ref{fig:donemodels}~\&~\ref{fig:donedem}).  

\begin{figure}
\includegraphics{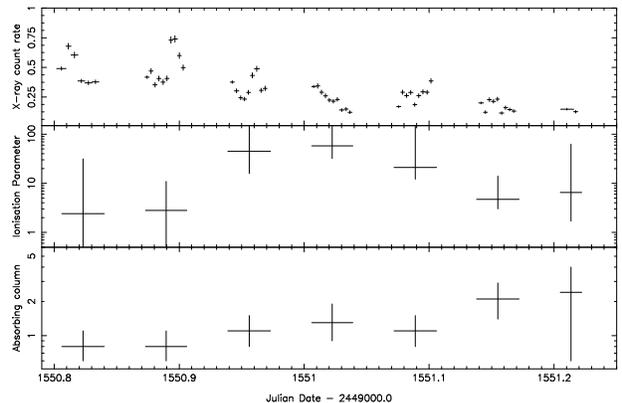}
\caption{The \asca X-ray light curve (top), the best-fitting DO97 model ionisation parameter ($\xi$ in units of L/nR$^2$, centre) and the best-fitting DO97 model absorbing column ($n_H, \times10^{22}$ atoms/cm$^2$, bottom), showing the spectral variations during the transition observation.  SIS and GIS spectra have been extracted from each \asca orbit, and fitted simultaneously with the DO97 model as described in the text.   Note that both the ionisation parameter and the absorbing column are plotted logarithmically.  Errors bars in all three panels are one sigma (68.3\% confidence level).}

\label{fig:variations}
\end{figure}

  To investigate the changing conditions during the transition observation, spectra from both the SIS and GIS has been temporally divided into seven spectra, with each spectra fitted with the DO97 model as before.  The variation of the \asca count rate, the DO97 best-fitting absorbing material ionisation parameter and absorbing $n_H$ are shown in Fig.~\ref{fig:variations}.
  Although there is the indication of variations in both the ionisation parameter and the absorbing column, the statistical uncertainties are too large to permit any firm conclusions to be drawn.  The temperature variations of the partially ionised absorber are unconstrained and so are not plotted in Fig.~\ref{fig:variations}.

\subsection{Flux Measurements}

Z~Cam has previously been observed with both \rosat (Wheatley \etal 1996a) and \exosat (Mukai \& Shiokawa, 1993). 
Our best-fitting DO97 model yields a 0.1-2.5\,keV flux of $7.7\times10^{-12}$ergs~cm$^{-2}$~s$^{-1}$ in the transition and $1.6\times10^{-12}$ergs~cm$^{-2}$~s$^{-1}$ during outburst. 
These values are a factor of 3 and 6 fold greater in quiescence and outburst respectively than during the \rosat observations (Wheatley \et, 1996a).
  Our transition 2-10\,keV flux of $2.5\times10^{-11}$ergs~cm$^{-2}$~s$^{-1}$ is 50\% greater than that measured by \exosat in quiescence (Wheatley \et, 1996a). 

\section{Discussion}

\subsection{Light Curves \& Hardness Ratio}

The \asca light curves of Z~Cam show that the X-ray count rate is greater during the transition to outburst than during outburst, supporting  other observations that show X-rays are suppressed during outburst.  
However, Fig.~\ref{fig:opticalxray} shows no spectral softening to the outburst level as the transition progresses.
  During the transition, the optical rise is expected to lead the fall in X-rays, allowing time for the outburst material to travel through to the inner disk.  As the material accretes onto the white dwarf surface the boundary layer becomes optically thick, thus suppressing the hard 
X-ray emission (Pringle 1977).
From Fig.~\ref{fig:opticalxray} this delay is 0.5-1.5~days, with most of the uncertainty originating from the determination of the beginning of the optical outburst.
Therefore, the beginning of the transition observation (from JD~2450550.79 to JD~2450551.02) is thought to have been made while the inner disk was still in a quiescent state.
The later part of the transition observation may have observed the inner disk of Z~Cam in true transition from the quiescent to the outburst state, although there is little difference in the spectra.  

\subsection{Temperature distribution}

  We observe high-temperature ($\approx$10\,keV) emission from the boundary layer in all our model fits, which is reduced during outburst and greater in the transition.
Fitting a discrete, two-temperature model to the transition and outburst observations simultaneously yields a remarkable fit with only the normalization of the higher temperature component varying between each state (see table~\ref{table:2ts}).

  Because the flux from the putative cooler component apparently does not vary, we investigate the possibility that the cooler emission may originate from the secondary star, which is expected to emit X-rays from a hot corona.
  The secondary star of Z~Cam is a dwarf of spectral type K7 (Ritter \& Kolb, 1998), which we assume to be close to the main sequence with a bolometric luminosity of $0.1L_{\odot}=3.8\times10^{32}$ erg s$^{-1}$.  The maximum coronal
emission from such a star is typically $10^{-3}L_{bol}$ (e.g. Pye \et 1994, Randich \et 1996), implying a maximum coronal luminosity for the secondary star in Z~Cam of approximately $4\times10^{29}$ erg s$^{-1}$.  In our two discrete temperature modelling, the lower temperature component luminosities are greater than this value ($\gs10^{30}$erg s$^{-1}$).  The difference between these two values are not significant enough to rule out the secondary star as the source of the lower temperature component. 

Since a continuous distribution of temperatures is more physically reasonable, we consider that this apparent cooler component is probably an artifact of an incorrectly modelled ionised absorber.
Therefore, we favour the continuous temperature distribution DO97 model which fits the data equally well.
The power-law emissivity function is much steeper during the spectrally-harder transition than during outburst, as demonstrated in Fig.~\ref{fig:opticalxray}.  During the transition the temperature distribution is more like that of a cooling flow, whereas during outburst there is an excess of cool gas.  This changing temperature distribution between the transition and outburst has also been observed in \asca observations of SS~Cyg (Done \& Osborne, 1997).

Although it is not possible to constrain the reflected continuum component, the fits require a fluorescent 6.4\,keV iron K line, with an equivalent width of $160^{+50}_{-60}$eV during the transition (F-statistic=34.4), and $400^{+120}_{-90}$eV during outburst (F-statistic of 6.7, probability$>$99.99\%).  Such a line is a natural consequence of reflection from cool gas.

\begin{figure}
\includegraphics{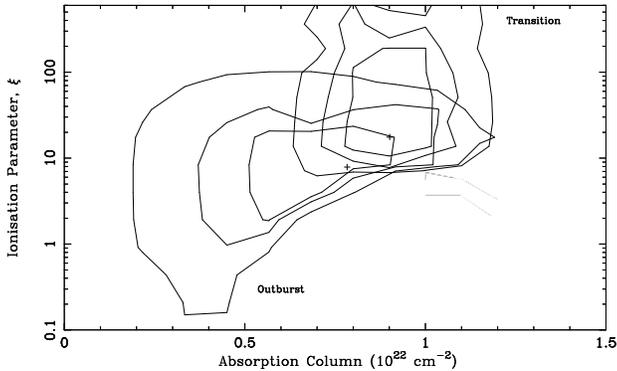}
\caption{Confidence limits on the ionisation parameter $\xi$ and hydrogen column absorption ($10^{22}$atoms cm$^{-2}$) from the DO97 model fits.  The contours represent 68\%, 95\% and 99\% confidence for the two parameters of interest.}
\label{fig:nhvsxi}
\end{figure}

\subsection{Absorption}
  The DO97 spectral fits suggest that there is
absorption beyond interstellar absorption by two orders of magnitude in both outburst and the transition.
The DO97 model fits suggests that the amount of ionised absorber remains approximately constant throughout both outburst and the transition, with $7.8^{+0.4}_{-0.3}\times10^{21}$ atoms cm$^{-2}$ in 
outburst, and $9.0^{+1.6}_{-1.5}\times10^{21}$ atoms cm$^{-2}$ in the transition.  
The level of ionisation of the absorbing material remains high in both observations, at $18^{+90}_{-8}$ in the transition observation to $8^{+67}_{-5}$ during outburst (errors are to the 90\% confidence level, see Fig.~\ref{fig:nhvsxi}), and both the ionisation parameter and the absorbing column remain high throughout the transition (see Fig.~\ref{fig:variations}).  The values of the ionisation parameter are consistent with those obtain through studies of SS~Cyg ($\xi=1.4^{+0.3}_{-0.3}$, Done \& Osborne 1997).

  Knigge \etal (1997) analysed the UV spectrum of Z~Cam during the same outburst, and derived the absorption column to be $\approx10^{22}$ atoms cm$^{-2}$ assuming a ionised, solar abundance plasma.  They suggested that the absorption in Z~Cam is due to a disk wind transition region above and below the accretion disk (c.f. Proga, Stone \& Drew, 1998).  
The \asca observations presented in this paper support this,
showing that the X-rays produced at the boundary layer were absorbed by the same amount of gas as 
the ultraviolet photons in both transition and outburst.  This suggests that the absorption is associated with vertical structure in the accretion disk, i.e. a clumpy disk wind, and is not associated with the boundary layer.

A disk wind is expected during outburst.  However, our analysis indicates that an absorbing wind may also be present early in the transition to outburst (see Fig.~\ref{fig:variations} and Fig.~\ref{fig:nhvsxi}).
Spectral modelling suggests that the ionisation state of the wind is high during both the outburst and during the transition.  This implies that either the disk wind starts in the earliest stages of outburst, or is present throughout the outburst cycle.

\section*{ Acknowledgments }

We thank Chris Done, Manabu Ishida, Hajime Inoue, Putra Mahasena and Chris Mauche for useful discussions, and also Chris Mauche for the calculation of the interstellar absorption.  
We thank the AAVSO for the optical light curves, which is based on observations submitted to the AAVSO by variable star observers worldwide.  We also thank the referee, Christian Knigge, for his helpful comments on improving the paper.
This research has made use of data obtained from the Leicester Database 
and Archive Service (LEDAS) at the Department 
of Physics and Astronomy, Leicester University, UK.
DSB acknowledges support of a PPARC studentship (UK) and a two month Monbusho fellowship (Japan).

\section*{References}
\normalsize

Anders E., \& Grevesse, N., 1989, Geochimica et Cosmochimica Acta, 53, 197\\

Arnaud K.A., 1996, Astronomical Data Analysis Software and Systems V, Eds. Jacoby G. and Barnes J., p17, ASP Conf. Series volume 101.\\ 

Cordova F.A., Chester T.J., Tuohy I.R., \& Garmire G.P., 1980, ApJ, 235, 163\\

Cordova F.A., Chester T.J., Mason K.O., Kahn, S.M., \& Garmire G.P., 1984, ApJ, 278, 739\\

Done C. \& Osborne J.P., 1997, MNRAS, 288, 649\\

Knigge C., Long K., Blair B., \& Wade R., 1997, ApJ, 476, 291\\

Liedahl D.A., Osterheld A.L., \& Goldstein W.H., 1995, ApJL, 438, 115\\

Mauche C.W., Raymond J.C., \& Cordova F.A., 1988, APJ, 335, 829\\

Mason K.O., Cordova F., Watson M.G., \& King A.R., 1988, MNRAS, 232, 779\\

Mewe R., Gronenschild E.H.B.M., and van den Oord G.H.J., 1985, A\&AS, 62, 197 \\

Mewe R., Lemen J.R., \& van den Oord G.H.J, 1986, A\&AS, 65, 511\\

Morrison R., \& McCammon D., 1983, ApJ, 270, 119\\

Mukai K. \& Shiokawa K., 1993, ApJ, 418, 863\\  

Mukai K., Wood J. H., Naylor T., Schlegel E. M., \& Swank J.H., 1997, ApJ, 475, 812\\

Patterson J., \& Raymond J.C., 1985a, APJ, 292, 535\\

Patterson J., \& Raymond J.C., 1985b, APJ, 292, 550\\

Pringle J.E., 1977, MNRAS, 178, 195\\

Pringle J.E., \& Savonije G.J., 1979, MNRAS, 187, 777\\

Pringle J.E., Bateson F.M., Hassal B.J.M., Heise J., Van der Woerd H., Holberg J.B., 
Polidan R.S., Van Amerongen S., Van Paradijs J., \& Verbunt F., 1987, MNRAS, 225, 73\\

Proga D., Stone J.M., \& Drew J.E., 1998, MNRAS, 295, 595\\

Pye J.P., Hodgkin S.T., Stern R.A. \& Stauffer J.R., 1994, MNRAS, 266, 798\\

Randich S., Schmitt J.H.M.M., Prosser C.F., \& Stauffer J.R., 1996, A\&A, 305, 785\\ 

Ritter H., \& Kolb U., 1998, A\&AS, 129, 83.\\ 

Shafter A.W., Ph.D. thesis, UCLA, 1983\\

Tanaka Y., Inoue H., \& Holt S., 1994, PASJ, 46, 37\\

Warner B, 1987, MNRAS, 227, 23\\

Wheatley P. J., van Teeseling A., Watson M. G., Verbunt F., \&
Pfeffermann E., 1996a, MNRAS, 283, 101\\

Wheatley P. J., Verbunt F., Belloni T., Watson M. G., Naylor T.,
Ishida M., Duck S.R., \& Pfeffermann E., 1996b, Astr.Astrophys., 307, 137.\\
  
\end{document}